# A Conclusive Experimentation Evidences that Mental States Follow Quantum Mechanics. Further Experimentation Indicates that in Mind States Bell Inequality Violation is Possible.

Elio Conte [1,2]

[1] *Department of Pharmacology and Human Physiology – TIRES – Center for Innovative Technologies for Signal Detection and Processing, University of Bari- Italy;*

[2] *School of Advanced International Studies for Applied Theoretical and Non Linear Methodologies of Physics, Bari, Italy;*

**Andrei Yuri Khrennikov** [3]

[3] *International Center for Mathematical Modeling in Physics and Cognitive Sciences, MSI, University of Växjö, S-35195, Sweden;*

**Orlando Todarello** [4]

[4] *Department of Neurological and Psychiatric Sciences, University of Bari, Italy;*

**Antonio Federici** [1]

**Joseph P. Zbilut** [5]

[5] *Department of Molecular Biophysics and Physiology, Rush University Medical Center, 1653 W, Congress, Chicago, IL 60612, USA.*

**Abstract:** In the first part of the paper we reach an experimental final confirmation that mental states follow quantum mechanics. In the second part further experimentation indicates that in mind states Bell inequality violation is possible.

## 1. Introduction

Human experience involves images, intentions, thoughts, beliefs. It consists of a content and the awareness of such content.

Consciousness is a system which observes itself. It evaluates itself being aware at the same time of doing so.

Let x, y, z be statements describing contents of various experiences. They are (atomic) content statements. Starting from such statements, other content statements may be formed by Boolean functions. $z = f(x, y,\ldots)$ are methalinguistic propositions that are not content statements or experience-describing statements themselves. Awareness statements a, b, c, …, self-referential or autoreferential, require autoreferential definitions as

$a_i = F_i(a_1,\ldots,a_n; x_1, x_2,\ldots,x_n)$, $(i = 1,2,\ldots,n)$ where $F_i$ are Boolean functions. As example $a = F(a,x)$ is the most simple definition of a single autoreferential statement $a$ [1]. $x = $ *the snow is white* ; $a = $ *I am aware of this*. We have here a logical self-reference mathematical model of conscious experience.

Consciousness represents the hard problem for excellence in our scientific, epistemological and philosophical knowledge [2].

If one day we would arrive identifying at all or in part the basic physical, scientific, philosophical principles or rules that act in our reality in order to determine consciousness and mind in humans, we would have moved in the way of a new great advance in our knowledge. This paper moves in such direction and perspective.

The present physical theory has not a definite apparatus to describe conscious systems. However, we cannot exclude that future generalizations of the present physical knowledge will be able to approach such basic problem. An indication arises from quantum mechanics. Quantum theory represents the most confirmed and celebrated theory of science. Started in 1927 by founder fathers as Bohr, Heisenberg, Schrödinger, and Pauli [3], it has revolutionized our understanding of the physical reality in both scientific and epistemological fields. It was introduced to describe the behaviour of atomic systems but subsequently its range of validity has turned out to be much wider including in particular some macroscopic phenomena like superconductivity or superfluidity. There is a salient and crucial feature for this theory. The conceptual structure and the axiomatic foundations of quantum theory repeatedly suggested from its advent and in the further eighty years of its elaboration that it has a profound link with mental entities and their dynamics. From its advent such theory was strongly debated but often also criticized just for its attitude to prospect a model of reality that results strongly linked to mental entities and their dynamics. In any manner standard formulation of quantum mechanics seemed to fix from its beginning the necessity to admit the unequivocal presence and role of mental properties to represent properties of the physical objects. This statement is true. We retain that it represents an important feature of the theory instead of its limit. However, there is the problem to correctly interpret such connection between quantum mechanics and mental properties in the sphere of our reality. It must be clear that one cannot have in mind a quantum physical reduction of mental processes. N. Bohr [4] borrowed the principle of complementarity, that is at the basis of quantum mechanics, from psychology. He was profoundly influenced from reading the "Principles of Psychology " by W. James [4]. However, N. Bohr never had in mind quantum-reductionism of mind entities. Starting with 1930, there was also an important correspondence between W. Pauli and C.G. Jung [5]. It culminated in the formulation of a theory called of mind-matter synchronization. Also in this case these founding fathers as Pauli and Jung were very distant to consider a quantum-reductionist perspective. V. Orlov [6] proposed to use quantum logic to describe functioning of brain, but also he did not look for reduction of mental processes to quantum physics. The correct way to frame the problem is not to attempt a quantum reduction of mental processes, the most profitable applications of quantum mechanics in cognitive sciences and psychology can be obtained not by any attempt of quantum physical reduction but giving experimental evidence that cognitive systems are very complex information systems, to which also some laws of quantum systems can be applied. Just the reaching of such objective would represent a very great advance in the domain of knowledge. In fact, starting with such experimental evidence, we could elaborate some future developments knowing this time the principles to use, the formal criteria to follow in order to approach with higher rigour the framing of the nature of mental entities and of their dynamics.

We retain that in this perspective we give here a first contribution since we give her for the first time experimental confirmation that mental states, at some stages of human perception and cognition, follow quantum mechanics. Thus for the first time, also if not under a reductionism perspective, we have the change to understand what are the principles and rules acting as counter part of human mind.

**2. The Theoretical Basis of the Experiment**.
To fully agree with the present paper, the reader must take care the following crucial point: quantum mechanics has its unique law of transformation of probability distribution. It is well known that the main feature of quantum probabilistic behaviour is the well known phenomenon of interference of probabilities. Such interference regime may be obtained only in quantum systems, e. g., in the celebrated two slit experiment that has been confirmed at any level of experimental investigation [7]. The interference gives the experimental basis of the superposition principle and this latter is the basis foundation of the physical and philosophical system of view that we call quantum mechanics. This is the essential peculiarity that we aim to investigate in the present paper. Recently, the problem of quantum probabilities was extended in the so called calculus of contextual probabilities [see in detail 8] .The essential feature of this elaboration is that by it we may be able to ascertain the presence of quantum like behaviour also in systems that exhibit context quantum like behaviour as physical, cognitive, social systems. We will not enter in the detail of the method here for brevity but all the features are given in the quoted literature [8]. The essence of the method is based on the following step. Let A and B be two dichotomic questions which can be asked to people

S with possible answers "yes (+) or not (-). In our case we consider A and B two mental quantum like observables of people S under investigation. We split the given ensemble S of humans in two sub ensembles U and V of equal numbers. To ensemble U we pose the question A with probability in answering, given respectively by $p(A=+)$ and $p(A=-)$, and $p(A=+)+p(A=-)=1$. We pose the question B immediately followed by the question A to the ensemble V. We will calculate conditional probabilities $p(A=+/B=+)$ and $p(A=+/B=-)$ and equivalent probabilities for the case $(A=-)$. Now we have reached a no eludible feature of such experiment. In such kind of transformation, by using the numerical results of the experimentation [8], we calculate

$$\cos \vartheta = \frac{p(A=+) - p(B=+)p(A=+/B=+) - p(B=-)p(A=+/B=-)}{2\sqrt{p(B=+)p(A=+/B=+)p(B=-)p(A=+/B=-)}} =$$

$$= \frac{\Delta p}{2\sqrt{p(B=+)p(A=+/B=+)p(B=-)p(A=+/B=-)}} \qquad (2.1)$$

If it results $\cos \vartheta \neq 0$ it will be certain that we are in presence of quantum like behaviour for mental states owing to the presence of interference terms for the calculated probabilities. In the case $\cos \vartheta = 0$ we will conclude that quantum mechanics is absent in the dynamic regime of our mental states. $\vartheta$ is obviously a well known angle of phase. In conclusion our experiment does not admit exceptions. By using the (2.1) it gives direct answer to the central problem to identify if quantum mechanics is involved or not in the regimes of mental states in humans. If we obtain $\cos \vartheta \neq 0$, the answer is positive while instead is negative if it results $\cos \vartheta = 0$.

We may proceed giving a quantum like framework of mental states. Let us remember that, according to Born's probability rule [3, 8], we have

$$P(A=\pm) = |\varphi(\pm)|^2 \qquad (2.2)$$

In the case in which the experiment confirms quantum mechanics in dynamics of mental states, as usually in standard theory, we can write a quantum-like wave function $\varphi_S(\pm)$ relative to the mental state $S$ of the population investigated, and it will be represented by the complex amplitude as for the first time elaborated in [8] and applied in our previous papers.

$$\varphi_S(x) = [P(B=+)P(A=x/B=+)]^{1/2} + e^{i\vartheta(x)}[P(B=-)P(A=x/B=-)]^{1/2} \text{ with } x=\pm \quad (2.3)$$

It is only necessary to outline the importance of future studies on cognition based possibly on the (2.3).

## 3. The Arrangement of the Experiment

Our experiment was based on the search of quantum behaviour in mental states during human perception and cognition of ambiguous figures.

On a general plane, it is known that brain organizes the sensory input into some representation of the given environment. Studies of perception indicate that the mental representation of a visual perceived object at any instant is unique also if we may be aware of the possible ambiguity of any given representation. The case is the Necker cube published by this author in 1832 [9] where we see the cube in one of two ways but only one of such representations is apparent at any time. We may be able to see the ambiguity of the design and even we may be able to switch wilfully between representations: we can be aware that multiple representations are possible but we can perceive them only one at time, that is seriatim.

Bistable perception arises whenever a stimulus can be thought in two different alternatives ways. In previous papers [8] we proposed to describe bistable perception with the formalism of a two quantum system. In our quantum like model of mental states we admit that an individual can potentially have multiple representations of a given choice situation, but can attend to only one representation at any given time. In this quantum mechanical framework we distinguish a potential and an actual or manifest state of consciousness. The state of the potential consciousness will be represented by a vector in Hilbert space. If we indicate as example a bi dimensional case with potential states $/1>$ and $/2>$, the potential state of consciousness will be given by
$$\psi = a/1> + b/2>. \qquad (3.1)$$
Here, $a$ and $b$ represent probability amplitudes so that $|a|^2$ will give the probability that the state of consciousness, represented by percept $/1>$, will be finally actualised or manifested during perception. Instead $|b|^2$ will represent the probability that state(percept) $/2>$ of consciousness will be actualised or manifested during perception. It will be $|a|^2 + |b|^2 = 1$.

**4. The Experiment Set Up.**
Generally speaking, the problem is to explain how, given multiple possibilities of representation, a particular representation can take place over our attention. In the case of Necker cube transitions between percepts may be possibly stochastic but in more complex mental and psychological situations some underlying factors may give the edge to one representation over another. Recent or repeated prior use of a representation may play role in advantaging one representation on the other. This is the reason to project the experiment carefully. Otherwise the study of ambiguous figures has intrigued and still is of valuable interest for psychologists and neuroscientists. A variety of theories has been published [10]. For the purposes of our experimentation we evaluated that two types of observers have been acknowledged by the experiments, fast observers having larger frequency of perspective reversals and slow observers whose frequency is lower. The persistent times staying one of the two percepts are usually in mean on the order of two seconds but may arrive also to about five seconds. To further confirm our quantum model with potential and actual states of consciousness, we have a further phenomenological datum. Subjects evidence uncertain time in states percepts in addition to perspective reversal. Uncertain times about 1 sec were experimentally ascertained in mean for speed subjects [11]. In conclusion, two kinds of times are identified during experiments with ambiguous figures: a time persisting one of the two possible percepts that we may call time persisting percept A, but it is similar to time persisting the percept B. Still we have a time of uncertain and neither one of the percepts nor another is certain for the subject, and this is in agreement with the previous quantum model on potential state of consciousness. Experiments have confirmed that the distribution of the persistent time in uncertain states is quite different from those corresponding to percepts A and B that result instead of the same behaviour and order. There are still two basic different approaches in studies of perception of ambiguous stimuli. One is the behavioural response to a stimulus based on psychological or mental processes. There are still two basic different approaches in studies of perception of ambiguous stimuli. One is the behavioural response to a stimulus based on psychological or mental processes. This is obtained using the frequency of reversals. The second approach looks instead to neural correlates of psychological processes triggered by stimuli.

We analyzed EEG recorded from parietal and frontal areas, focusing on gamma band phase synchronization between these two areas. A recent neuroimaging study using functional magnetic resonance imaging (fMRI) has suggested that conscious detection of visual changes relies on both parietal and frontal areas [12]. The same areas also show activation in perceptual switching experiments [12]. These areas, therefore, seem to play an important role in detecting changes in our perception, whether they are caused externally or internally [12]. The temporal characteristics of the relationship between these two areas are not well understood. However, it has been proposed that transient synchronization of oscillatory activity, in particular in the gamma band of the EEG, plays an important role in building coalitions between areas [12]. Moreover transient gamma band synchronization induced by sensory stimulation is considered important for perceptual feature binding of distributed representations [12]. For this reason, transient gamma band synchronization is important in our investigation. The time scale of 3 sec results from a state of being conscious and that this time scale is also significant for cognitive processes beyond the bistable perception of ambiguous stimuli. Experiments have been performed [12] concerning capabilities of discriminating and sequencing temporally separate perceptual events. Also for $\Delta t \phi 30 ms$, two different individual events are clearly separable and their sequence can be correctly assigned. In conclusion these are the two cognitive time scales that one should take in consideration in experimentation on perception and cognition of ambiguous figures.

Our experiment was planed on an accurate consideration of the data previously obtained in literature. Its architecture was based on the analysis of the (2.1) with given ambiguous figures A and B.
Starting with 2003 we performed our previous three experiments [8] of this kind and they were based on ninety eights subjects. In the present experimentation we investigated a group of other twenty six subjects. In substance, this last experimentation concludes our cycles of verification on this team after two years of investigation and one hundred and twenty four subjects examined. It must be clear that for The last twenty six subjects that we engaged in such final experimentation, we used more restrictive conditions of experimentation in the sense that we used different tests of ambiguous percepts, and we made any effort to use ambiguous figures giving to the subject an immediate effect of visual ambiguity followed soon after by a direct and unique perception of the subject. The high speed in inducing ambiguity followed from a rapid selection of percept gives the best experimental conditions to examine our quantum model of consciousness. In this sense we retain, therefore, that the results of the present investigation must be considered more definitive respect to the other results that we performed previously. In fact ,only to repeat we remember that, according to our quantum model as previously given in (2.1), we admit that an individual can potentially have multiple representations of a given choice situation, but he can attend to only one representation at any given time. Strong and immediate ambiguity as induced in the present case by test A and B, would consequently induce the subject to suspend his potential consciousness state soon after followed from actualised or manifest state of his consciousness. In conclusion, by this last phase of experimentation, we retain to have performed any effort to guarantee the best final conditions in verifying if quantum mechanics enters in the dynamics of mental states during perception of ambiguous figures. All the subjects were selected with about equal distribution of females and males, aged between 19 and 22 years. All had normal or corrected-to-normal vision. All they were divided by random selection in two groups (1) and (2) of thirteen subjects. The group (1) was subjected to test A (Fig.A) while the group (2) was subjected to Test B and soon after (about 800 msec after choice for test B) to test A. In all the cases the posed question to the subject was to answer what animal he was thinking after certain identification of ambiguity when he had in vision the figure. It has been shown that perception and cognition in ambiguous figures is influenced by visual angle [13]. Therefore a constant visual angle $V = 2 arctg(S/2D) = 0.33\ rad$. was used by us in our experimentation with $S$ object's frontal linear size and $D$ distance from the center of the eyes for all the subjects. Each observer was seated at a table with a monitor and computer, he was told to look binocularly at the figure, no fixation point was given, and it was adopted a passive attitude towards the figure without any effort to hold one percept or to favour reversals. The observer was requested to stop by pressing a key at the computer when he was aware to have thought one percept and only one after direct verification of the existing ambiguity in the figure. The ambiguous figures were placed in front of the eyes of the observer at a distance of 60 cm, and illuminated by a lamp of 60 W located above and behind the observer's had. The experimental room was kept under daylight illumination. The constant visual angle was realized for each subject using an S object's frontal linear size of about 26 cm for the figure on the monitor.

## 5. Results of the Experiment
The results of the present experimentation confirmed that mental states follow quantum mechanics during perception and cognition of ambiguous figures. They still confirm the results that on the same subject we had previously obtained with some different experimental arrangements. So we retain that by such our final experimentation we have reached a final conclusion on this subject On the basis of the (3.1) we may confirm that mental states follow quantum mechanics during perception and cognitive performance of human brain for ambiguous percepts and cognition. The engaged quantum model indicates that human beings can potentially have multiple representations of a given choice situation but can attend to only one representation at any given time via a quantum mechanical jump. In the quantum mechanical framework that we propose, we distinguish a potential and an actual or manifest state of consciousness. The state of the potential consciousness will be represented by a vector in Hilbert space. If we indicate as example a bi dimensional case with states indicated by $/1>$ and by $/2>$, the potential state of consciousness will be given by $\psi = a/1> + b/2>$. In this case the (2.1) must be verified. The results that we obtained, are reported are reported in Table1. It is seen that we obtain $\cos\vartheta(+) = -0.307$, and thus $\vartheta(+) = 1.882$. This result unequivocally confirms that mental states follow a quantum mechanical regime in presence of perception and cognition of ambiguous figures. In this manner, according to the (2.3) we may also write the quantum like wave function $\varphi_S(+)$ of the mental state $S$ of the group of human beings investigated, and it results to be given in the following manner (see the 2.3):

$\varphi(+) = [0.615 \times 0.750]^{1/2} + e^{i\vartheta(+)}[0.385 \times 0.800]^{1/2} \approx 0.679 + 0.554\,i$ for mental variable $A$ with assumed value +.

Analogous procedure we could use to calculate $\varphi(-)$ that is relative to the mental variable $A$ with assumed valued $A = -$.

In conclusion, in this paper we have reached two important results: we have given a rather definitive question that quantum mechanics enters in the dynamics of mental states during to perception and cognition of human beings of ambiguous features. We have also given body to some fundamental quantum mechanical elaboration as in particular the calculation of quantum like wave function for mental states. It cannot be excluded that this result also promises in perspective important advances on the plane of interpretation of dynamics of mind and its functions also in cases of normal and or subjects with pathologies.

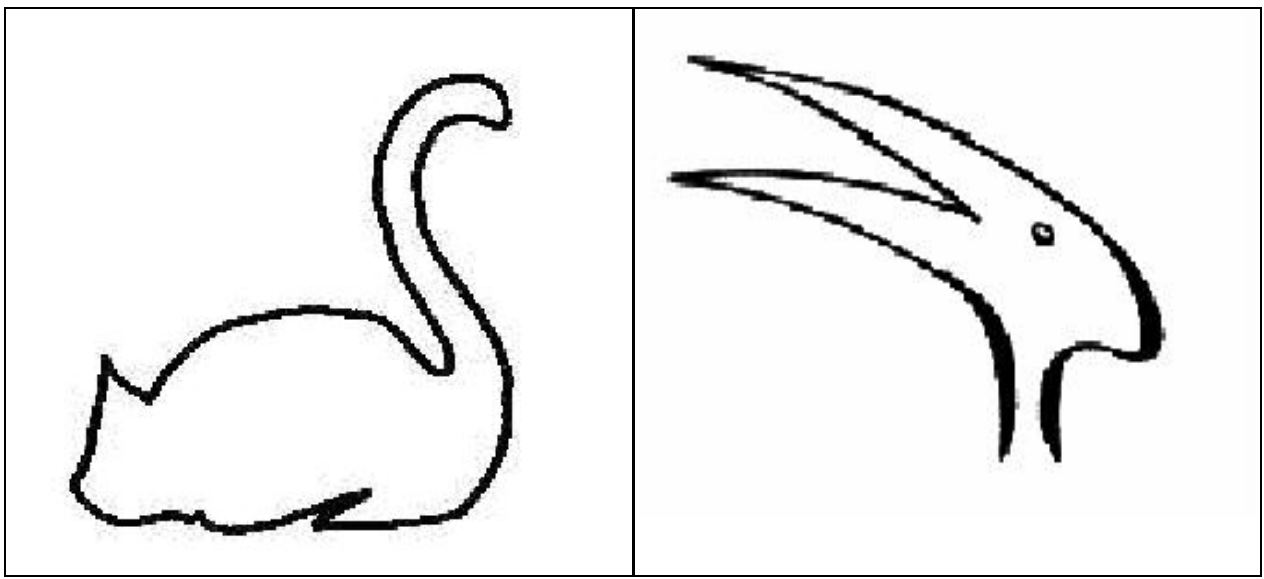

**Figure A**                                                **Figure B**

**Tab. 1**

|  | Test A | | | Test B | | A/B | |
|---|---|---|---|---|---|---|---|
| Subject | CAT (+) | SWAN (-) | | STORK (+) | RABBIT (-) | CAT (+) | SWAN (-) |
| 1 |  | * | | * |  | * |  |
| 2 |  | * | | * |  |  | * |
| 3 | * |  | | * |  | * |  |
| 4 | * |  | | * |  | * |  |
| 5 | * |  | |  | * | * |  |
| 6 | * |  | | * |  |  | * |
| 7 | * |  | |  | * | * |  |
| 8 |  | * | | * |  | * |  |
| 9 | * |  | |  | * | * |  |
| 10 |  | * | | * |  | * |  |
| 11 |  | * | |  | * |  | * |
| 12 |  | * | | * |  | * |  |
| 13 | * |  | |  | * | * |  |
|  | p(A=+) | p(A= -) | | p(B=+) | p(B= -) | p(A=+/B=+) | p(A=+/B= -) |
|  | **0.538** | **0.462** | | **0.615** | **0.385** | **0.750** | **0.800** |
| **Δp=-0.231** |  |  | | **cos θ=-0.307** |  | **θ=1.882** |  |

**Tab. 2**

### Statistical analysis: Student's t-test

|  | I | II | III | IV | V | VI | VII |
|---|---|---|---|---|---|---|---|
| **Experiments** | p(A=+) | p(A= -) | p(B=+) | p(B= -) | p(A=+/B=+) | p(A=+/B= -) | p(B=+)p(A=+/B=+)+p(B=-)p(A=+/B=-) |
| **1** | 0.6923 | 0.3077 | 0.9259 | 0.0741 | 0.6800 | 0.5000 | 0.6667 |
| **2** | 0.5714 | 0.4286 | 1.0000 | 0.0000 | 0.7000 | 0.0000 | 0.7000 |
| **3** | 0.4545 | 0.5455 | 0.7000 | 0.3000 | 0.4286 | 1.0000 | 0.6000 |
| **4** | 0.5380 | 0.4620 | 0.6150 | 0.3850 | 0.7500 | 0.8000 | 0.7692 |
| **Mean Value** | 0.5641 | 0.4360 | 0.8102 | 0.1898 | 0.6397 | 0.5750 | 0.6840 |
| **St. Dev.** | 0.0986 | 0.0986 | 0.1823 | 0.1823 | 0.1437 | 0.4349 | 0.0704 |

number of examined subjects n=124.
t-test results: P value=0.0951 (significance, P> 90%)
t=1.979

**We may now add the following integration of the previous results. They are summarised in Tab.3.**

The first time we group all the 124 subjects that we employed in all the experimentation and, considering it as only one group, independently of the nature of Tests A and B that we used, we proceed to the calculation of $\cos\vartheta$ and thus of $\vartheta$ for the only group of subjects. As seen by Tab.3 we obtain the value $\vartheta = 1.6942$. In the second case, instead, we considered still the group of the four experiments that we performed and on this basis we calculated $\vartheta = 1.722367 \pm 0.173423$ (For significance we add also the t-test value that was obtained in this case).

It seems to us that the result that we obtain is of valuable interest. In brief, we confirm first of all that mental states exhibit quantum like interference of probabilities in Human Subjects during their perception and cognition in perspective reversals of ambiguous visual patterns, but the most salient result seems that the quantum like regime of mental states happen with an invariant as phase angle characterizing quantum interference of probabilities. In the case of examining the whole group of subjects we had in fact $\vartheta = 1.6942$ while in the case of the four considered experiments separately, we had $\vartheta = 1.722367 \pm 0.173423$, that is very similar to the previous one. In conclusion, quantum like regime of mental states could be regulated by a presently unknown mechanism signed by an invariance in the phase angle expressing quantum like interference of probabilities in all normal subjects as those we investigated.

**Tab.3**

**Statistical analysis: Student's t-test. Subdivision within four performed experiments**

| Experiments | I<br>p(A=+) | II<br>p(A= -) | III<br>p(B=+) | IV<br>p(B= -) | V<br>p(A=+/B=+) | VI<br>p(A=+/B= -) | VII<br>p(B=+)p(A=+/B=+)+p(B=-)p(A=+/B=-) | cos(theta) | theta |
|---|---|---|---|---|---|---|---|---|---|
| 1 | 0.6923 | 0.3077 | 0.9259 | 0.0741 | 0.6800 | 0.5000 | 0.6667 | **0.0839** | **1.4867** |
| 2 | 0.5714 | 0.4286 | 1.0000 | 0.0000 | 0.7000 | 0.0000 | 0.7000 | | |
| 3 | 0.4545 | 0.5455 | 0.7000 | 0.3000 | 0.4286 | 1.0000 | 0.6000 | **-0.2425** | **1.8157** |
| 4 | 0.5380 | 0.4620 | 0.6150 | 0.3850 | 0.7500 | 0.8000 | 0.7692 | **-0.3067** | **1.8825** |
| Mean Value | 0.5641 | 0.4360 | 0.8102 | 0.1898 | 0.6397 | 0.5750 | 0.6840 | **-0.1331** | **1.7043** |
| St. Dev. | 0.0986 | 0.0986 | 0.1823 | 0.1823 | 0.1437 | 0.4349 | 0.0704 | | |

number of examined subjects n=124.
t-test results: P value=0.0951 (significance, P> 90%) t=1.979

**Results obtained when considering one experiment for the whole set of 124 subjects**

| p(A=+) | p(A= -) | p(B=+) | p(B= -) | p(A=+/B=+) | p(A=+/B= -) | p(B=+)p(A=+/B=+)+p(B=-)p(A=+/B=-) | cos(theta) | theta |
|---|---|---|---|---|---|---|---|---|
| 0.5968 | 0.4032 | 0.8387 | 0.1613 | 0.6346 | 0.8000 | 0.6613 | -0.1231 | 1.6942 |

**Acknowledgments**

One of us, E.C., gratefully acknowledges Prof. Matti Pittkanen for his valuable comments and criticism and for his important previous work entitled Experimental support for binocular rivalry as a quantum phenomenon that may be found in http://matpitka.blogspot.com/2007/10/experimental-support-for-binocular.html.

**PART II: On the possible Bell inequality violation in mental states**

1. **The Theoretical Framework**.

Bell's theorem was published in a fundamental paper of physics in 1964 [1]. In 1975 Stapp defined Bell's theorem "The most profound discovery of science" [2]. We have to observe that He speaks here respect to science and not only respect to physics. Let us remember what the theorem states: it shows that the predictions of quantum mechanics are not intituitive and they relate the most fundamental issues of our physical, physolosophical, epistemological and onthological reality. It enphasizes that no physical theory of local hidden variables may ever reproduces all of the predictions of quantum mechanics. It is the most famous legacy that we encounter in physics.

Of course it is well known that Einstein was critical respect to standard interpretation of quantum mechanics. The celebrated EPR paper [3] showed that the standard interpretation of such theory, implies either its incompleteness or "spooky action-at-a-distance". Einstein wanted to get rid of the "action-at-a-distance" by assuming incompleteness of quantum mechanics and introducing "local hidden variables." Bell's theorem, published in 1964, is considered to prove that it is possible to construct experiments in which it is impossible for any kind of interpretation based on "local hidden variables" to give the same predictions as quantum mechanics, providing a means of testing whether "action-at-a-distance" actually occurs. A lot of experiments have been conducted in physics confirming fully the validity of quantum heory [4].

Rather recently, on the theoretical plane, analysis has been performed of probabilistic assumptions of Bell's formulation. One of us [5], Andrei Khrennikov, has emphasized that J. Bell wrote about probability withouth to specify the concrete axiomatics of probability theory. His analysis shows that Bell did not apply the classical probability model, that, as it is well known, is based on Kolmogorov model to describe classical physical framework.In substance he introduced his own probabilistic model and compared it with quantum mechanics.The crucial point for the present paper is that he did not payed attention to conditional probabilities. In detail, [5], in this model it was shown that the conditional probabilities in Bell elaboration cannot be defined by classical Bayes formula. In [5] it was used the approach based on Bell- type inequalities in the conventional approach of the Kolmogorov model. Khrennikov showed an analog of Wigner inequality [6] for conditional probabilities, and, on this basis, evidenced in detail that the predictions of the conventional and quantum probability models must disagree.

Thia result represents an important advance because, in addition to the important theoeretical elucidation that it reaches, on its basis we become finally able to perform experiments in various fields having in this manner the possibility to discuss the possible violation of Bell inquality in very different fields of interest. Our aim is confined to investigation of mental states brain dynamics, and in paricular to perception and cognition of human beings. We will discuss in detail such featues of the argument within few time, but before let us add some further important considerations.

In particular in this moment, but the question was in some manner signed fom the starting of the theory, quantum mechanics suffers of a net dichotomy. We have a group of researchers that strongly is inclined to accept that quantum mechanics is a theory that speaks only about physical systems. In this manner it excludes, as example, the possibility to analyze cognitive systems by this theory and its basic fundations. On the contrary, there is another current of research that in principle is available to admit a larger field of pertinence of quantum mechanics, that is to say that it should not pertain uniquely to only physical systems but such enlarged field of reference could regard in particular brain dynamics, and, in particular, the sphere of the mental states in human beings. We retain to have given recently a net and definitive answer to this fundamental question. Accumulating results for about more than five years of experimental research, we have shown in a conclusive paper that that mental states follow quantum mechanics during perception and cognition of ambiguous figures in a very large group of human subjects. The results have been published by us in arXiv: 0802.1835 [7] and references therein.The mentioned papers contain all the elements to follow in detail the articulated experimentation that was performed.

Some points must be outlined respect to such results:

1) First of all, they confirm that quantum mechanics is an elaborated theory that does not confines itself to analysis of physical systems only as admitted by some groups of research.
2) For the first time we identify a physical theory that shows to have direct and fundamental relevance in the sphere of mental and brain dynamics with particular relevance during human perception and cognition. This is a result that opens new perspectives in the sphere of scientific knowledge since it gives a decisive contributin to the important problem to undertsand human consciousness indicating for the first time the way in which we may conceive and explore it in formal terms based on a physical theory. It is rather obvious to expect that , having identified for the first time the formal theory to be followed in treating mental states and consciousnes, new important perspectives are open with particular relevance for the corresponding correlates at the level of psychological but also neurophysiological levels. Recently [8] studies were conducted by various authors analysing in detail the electrophysiological correlates of perceptual reversals in the case of three different types of multistable images as the modified Rubin'sface/vase, the Necker cube and Lemmo's cheetahs. The results seem to support a model of multistable perception in wich changes in early partial attention modulate perceptual reversals.The link of this model with a possible quantum mechanical framework seems rather evident.
3) There is finally an onthological consideration. It is significant for the manner in which our results must be considered. To this purpose, any reductionistic approach must be excluded. Quantum mechanics must not be intended as the theory to which we reduce the functioning of the brain as many authors outlined in the past [9]. Brain dynamics represents the most extensive system holding on a very complex dynamics, and it deserves to be analized and undestood under different profiles including also, and in particular, some approaches of physiscs as chaos theory. Our results simply indicate that the principles, the epistemological and onthological foundations of quantum mechanics certainly enter in the so articulated description and analysis of consciousnes and of mental states of human beings. Quantum mechanics certainly enters in brain dynamics but it is not the only theory by which we may have the presumption to explain brain complexity. This is the most promising feature of our recent discovery**:** to have ascertained with a net degree of accuracy the real role of quantum theory in brain dynamics, but any reductionistic approach is excluded by us.

In order to support our thesis, we would now perform a step on. It would be to analyze mental states of a gropup of human subjects verifying directly the possibility and the conditions under which a violation of Bell inquality may turn out in the analysis of mental states of subjects during their perception and cognition of ambigous figures.

To this purpose in the present paper we have performed a first experiment on Bell inequality, arriving to establish the conditions in which it may be violated.

We will indicate here the results that we have obtained and we will indicate also the neurophysiological basis that we must follow in order to obtain Bell violation in perception and cognition of human during their perception of ambigous figures.

**2. The Theoretical basis of the Experiment**

As previously said, the aim of our program of experimentation was to analyse quantum-like behaviour of mental states by the analysis of Bell's inequality.
We followed step by step the previous results obtained in [5] to which the reader is re sent for the relative deepening. Let us start considering the well known Wigner inequality [6] that this author evidenced for Bell inequality [5].
Let $\mathbf{a}, \mathbf{b}, \mathbf{c} = \pm 1$ be arbitrary dichotomous random variables on a single Kolmogorov space S. Then the following inequality holds true
$$p(a = +1, b = +1) + p(c = +1, b = -1) \geq p(a = +1, c = +1) \tag{1.1}$$
The proof of this theorem and a detailed discussion on the subject may be found in [5,7].
We recall now that conditional probabilities in the Kolmogorov model are defined by using Bayes' formula and thus writing
$$p(a = a_1 / b = b_1) = \frac{p(a = a_1, b = b_1)}{p(b = b_1)} \tag{1.2}$$
The aim of the test that we perform is easily explained :
**A family of observables which does not permit such statistical realistic description will be called to follow a quantum like behaviour**.
Using the (1.2) in the (1.1), we obtain that
$$p(a = +1 / b = +1) + p(c = +1 / b = +1) \geq p(a = +1 / c = +1) \tag{1.3}$$
**that is the analogue of Bell's inequality for conditional probabilities**. If this inequality is violated in experiments by mental states we will conclude that in this particular experimental, the mind states behave quantum like violating Bell inequality with respect to questions a, b, and c.

**3. The Description of the Experiment**

A group S of 144 subjects was divide into two subgroups U of 72 subjects and V of 72 subjects. To the group U the test B was given with possible answers $\mathbf{B}^+$ and $\mathbf{B}^-$. To subject answering $\mathbf{B}^+$, it was subsequently given the test $\mathbf{A}$ with possible answers $\mathbf{A}^+$ and $\mathbf{A}^-$ respectively. To subjects answering $\mathbf{B}^-$, it was given instead the test $\mathbf{C}$ with possible answers $\mathbf{C}^+$ and $\mathbf{C}^-$. To subjects pertaining the subgroup V, it was given test instead the test **C** with **C**$^+$ and **C**$^-$ as possible answers. To subjects answering **C**$^+$, it was finally given the test **A** with possible answers **A**$^+$ and **A**$^-$. The scheme of the experiment is given in Fig.1 and it follows rigorously the (1.3).

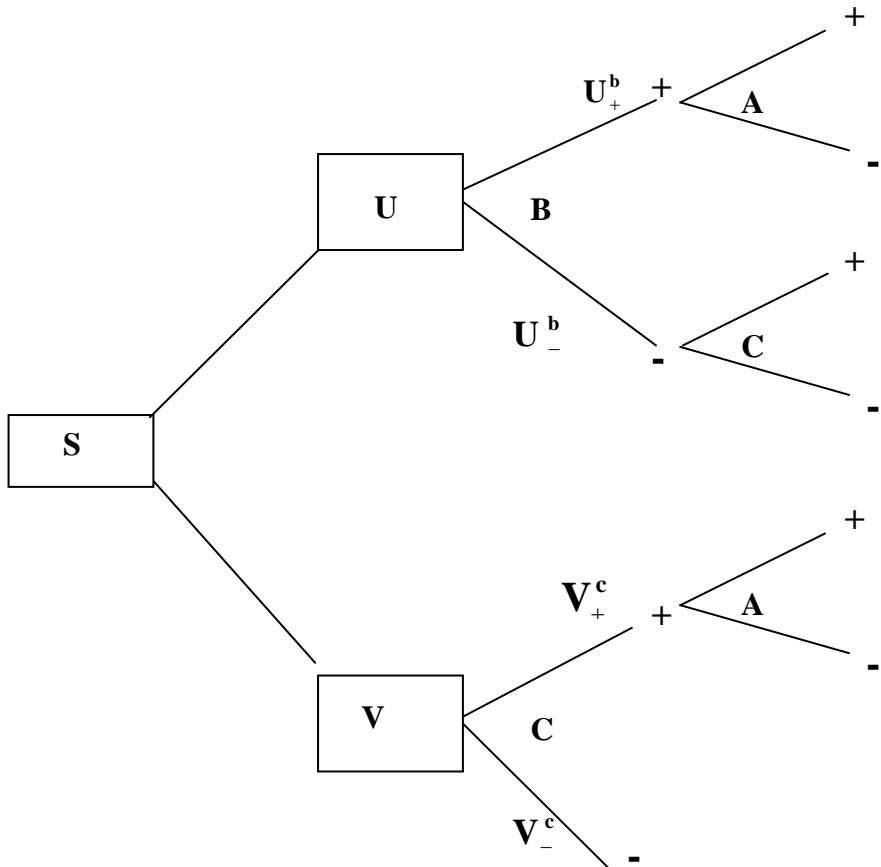

**Figure 1. The scheme of the performed experiment**.

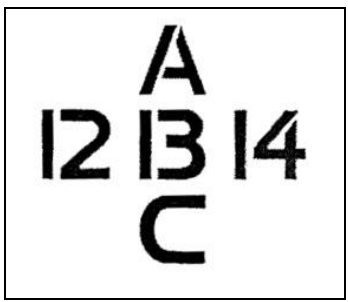  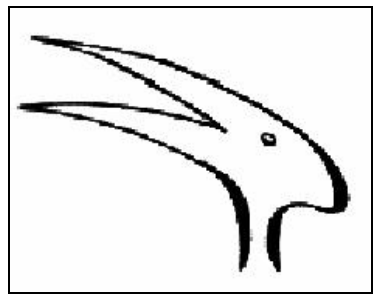  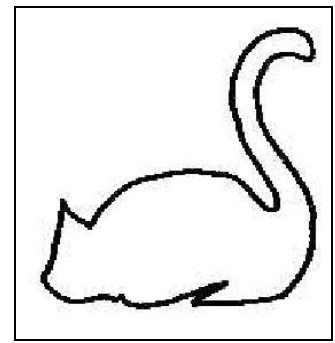

**Test B**  **Test C**  **Test A**

Let us summarize the data and the results that were obtained

Investigation group of 144 subjects:
U = 72 subjects
V = 72 subjects

$U^b_+$ = 37 subjects    ;    $V^c_+$ = 45 subjects

$U_-^b$ = 35 subjects    ;    $V_-^c$ = 27 subjects

$$p(a=+/b=+) = \frac{22}{37} = 0.5946 \quad ; \quad p(c=+/b=-) = \frac{23}{35} = 0.6571 \tag{1.4}$$

$$p(a=+/c=+) = \frac{24}{45} = 0.5333$$

$$p(a=+/b=+) + p(c=+/b=-) \geq p(a=+/c=+) \tag{1.5}$$
0.5946 + 0.6571 ≥ 0.5333

**The result is that Bell inequality was not violated.**

Note, however, that in the hypothesis of the theorem shown in [5] we must have that

$$p(b=+) = p(b=-) = \frac{1}{2} \quad ; \quad p(c=+) = p(c=-) = \frac{1}{2}.$$

**Instead we had asymmetrical results:**
p(b=+) = 0.514    p(b=-) = 0.486   p(c=+) = 0.625    p(c=-) = 0.375
We had to return to verification of old Wigner inequality, that is
p (a=+1,b=+1)+p(c=+1,b=-1)>=p(a=+1,c=+1).
Since in a Kolmogorov model we have

$$p(a=+/b=+) = \frac{p(a=+;b=+)}{p(b=+)} \quad ; \quad p(c=+/b=-) = \frac{p(c=+;b=-)}{p(b=-)} \quad ;$$

$$p(a=+/c=+) = \frac{p(a=+;c=+)}{p(c=+)}$$

we calculated this time hat

p(a=+,b=+)=p(a=+/b=+) p(b=+) ; p(c=+,b=-)=p(c=+/b=-) p(b=-) ; p(a=+,c=+)=p(a=+/c=+) p(c=+) .
Consequently we had
p (a=+;b=+)=0.5946 x0.514 =0.30562 ; p(c=+,b=-)=0.6571 x 0.486=0.31935;
p(a=+,c=+)=0.5333 x 0.625 =0.33331
that is to say
p (a=+1,b=+1)+p(c=+1,b=-1) ≥ p(a=+1,c=+1)  as 0.30562+0.31935 ≥ 0.33331 .     (1.6)
In conclusion Bell inequality resulted not violated in such experimental stage of mental states behaviour during perception and cognition of ambiguous figures.

**4. Refinement of the previous calculations.**

We must give now some comments of valuable importance.
   a.    First of all we must outline that such results do not invalidate the results that recently we obtained in ref [7] where we showed that mental states follow quantum mechanics during perception and cognition of ambiguous figures. Inequalities of Bell kind by themselves have nothing to do with quantum theory. As correctly outlined by J.H. Eberly [10], contexts as different as downhill skiers [11] and laundered socks [12] were used to demonstrate this. This is the first important point to outline.
   b.    There is another important feature that was recently evidenced by Pitkänen [13]. He outlined that if the failure of inequality does not occur, this does not of course mean that the system is classical but only that the quantal effects, in the proper version that we experienced, were not large enough.
   c.    Finally, according to [13], the Bell inequality that we have used in the present experimentation suggests that the questions we posed to human beings, and consisting in

Tests A, B, C, previously seen, must be considered to be analogous to spins in spin pair of spin singlet states in an external magnetic field and determining a quantization axis. In other terms [13], the failure of Bell inequality is expected if the ambiguous figures were used by us in a manner that produced bistable percepts differing enough.

d. We arrive in this manner to the crucial conclusion of the present paper. In quantum theory we have, as usually admitted, quantum observables. They are physical quantum observables. Owing to the results the we obtained recently [7], and showing, we repeat, that mind states follow quantum mechanics during perception and cognition of ambiguous figures, we must conclude under a very new perspective : we have also mental observables, that as it was recently shown by us, follow quantum mechanics at least during perception and cognition of ambiguous figures. These mental observables are represented by the three abstract set $(e_1, e_2, e_3)$ of the algebraic structure that we have used several times in our papers [14]. They may pertain to the sphere of mind activity and logic as well as also to the physical sphere as it is the case of physical particles where the projections of spins in three perpendicular reference directions are represented by the well known Pauli matrices, $S_1 = \frac{1}{2}he_1$, $S_2 = \frac{1}{2}he_2$, $S_3 = \frac{1}{2}he_3$. Also the projections of the magnetic moment are given by physical observables as $\mu e_1, \mu e_2, \mu e_3$. However, the $e_i$ $(i = 1,2,3)$, considered as algebraic abstract elements in the manner discussed by us in a number of previous papers [14], represent mind-logical observables in a quantum like version of the theory.

e. Finally, if in the present case of the experimentation, given in figure 1, we consider the tests A, B, C, as expressed by the algebraic sets $(e_1, e_2, e_3)$, we arrive to find cases in which Bell inequality may be violated and we find also the neurophysiological correlates to such experimental condition.

f. Finally, we intend to evaluate here still the hypothesis to consider the algebraic set $(e_1, e_2, e_3)$ as mental logic observables. It is not new here. Yuri Orlov, formulating his well known wave logic, in 1982, in an important paper entitled "The Wave Logic of Consciousness: A Hypothesis" at pag. 41 wrote textually [15] :

*According to my paper presented at the 5$^{th}$ Congress on Philosophy, Methodology of Science and Mathematical Logic, every atomic proposition of classical logic can be represented by a diagonal operator-the third component of the Pauli algebra $e_3$ :*

$$e_3 = \begin{pmatrix} 1 & 0 \\ 0 & -1 \end{pmatrix} \quad , \quad e_3^2 = 1 \quad , e_3^* = 1$$

This so authoritative affirmation represents the most convincing support to our positions.
It remains now to show how Bell inequality could be violated in our experiment of Figure 1, based on mental states. Let us assume the Tests A, B, C, corresponding to mental observables, are given in the following manner:

Test A $= \cos\vartheta_1 e_3 + sen\vartheta_1 e_1$ ; Test B $= \cos\vartheta_2 e_3 + sen\vartheta_2 e_1$ ; Test C $= \cos\vartheta_3 e_3 + sen\vartheta_3 e_1$.

This may be now written on the basis of the previous positions that we introduced.
We have $AB \neq BA$, $AC \neq CA$, $BC \neq CB$

where $AB \equiv$ Test A Test B $= \cos(\vartheta_1 - \vartheta_2) - isen(\vartheta_1 - \vartheta_2)e_2$
and $BA \equiv$ Test B Test A $= \cos(\vartheta_1 - \vartheta_2) + isen(\vartheta_1 - \vartheta_2)e_2$
where $AC \equiv$ Test A Test C $= \cos(\vartheta_1 - \vartheta_3) - isen(\vartheta_1 - \vartheta_3)e_2$
and $CA \equiv$ Test C Test A $= \cos(\vartheta_1 - \vartheta_3) + isen(\vartheta_1 - \vartheta_3)e_2$ (1.7)
where $BC \equiv$ Test B Test C $= \cos(\vartheta_2 - \vartheta_3) - isen(\vartheta_2 - \vartheta_3)e_2$
and $CB \equiv$ Test C Test B $= \cos(\vartheta_2 - \vartheta_3) + isen(\vartheta_2 - \vartheta_3)e_2$.

We know that

$$p(a=\alpha_i / b=\beta_i) = |<\varphi_i^a, \varphi_i^b>|^2$$

being $\{\varphi_i^a\}$ and $\{\varphi_i^b\}$ normalized set of eigenvectors, respectively.
We have

$$\sigma(\vartheta)\varphi_+(\vartheta) = \varphi_+(\vartheta) \quad \text{with} \quad \varphi_+(\vartheta) = (\cos\frac{\vartheta}{2}, sen\frac{\vartheta}{2}) \quad \text{and}$$

$$\sigma(\vartheta)\varphi_-(\vartheta) = -\varphi_-(\vartheta) \quad \text{with} \quad \varphi_-(\vartheta) = (-sen\frac{\vartheta}{2}, \cos\frac{\vartheta}{2})$$

In conclusion, instead of the (1.3) and the (1.5) we obtain now that

$$p(a=+/b=+) = \cos^2\frac{\vartheta_1 - \vartheta_2}{2} \;;\; p(c=+/b=-) = sen^2\frac{\vartheta_3 - \vartheta_2}{2}, \text{ and finally}$$

$$p(a=+1, c=+1) = \cos^2\frac{\vartheta_1 - \vartheta_3}{2} \tag{1.8}$$

In conclusion, the (1.5), before non violated, now becomes

$$\cos^2\frac{\vartheta_1 - \vartheta_2}{2} + sen^2\frac{\vartheta_3 - \vartheta_2}{2} \geq \cos^2\frac{\vartheta_1 - \vartheta_3}{2} \tag{1.9}$$

that represents the central equation of our paper, for details see in particular ref [5 and 6].
According to [5,6], a simple solution of the (1.9) may be given assuming $\vartheta_1 = 0, \vartheta_2 = 6\vartheta, \vartheta_3 = 2\vartheta$.
Instead of the (1.9), one obtains that

$$\cos^2 3\vartheta + sen^2 2\vartheta \geq \cos^2 \vartheta \tag{1.10}$$

In Figure2 we give in red the sum of the two terms on the left and in blue the term on the right of the inequality (1.10).
As expected, we see that we have regions of $\vartheta$ in which Bell's inequality for mental states may be violated.

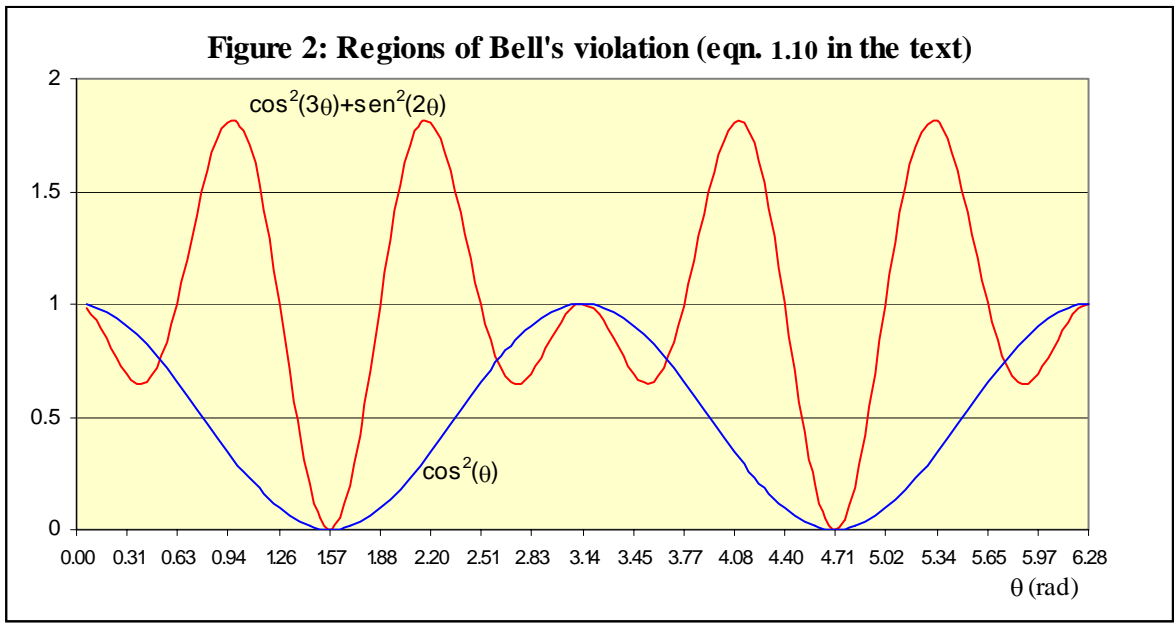

Figure 2: Regions of Bell's violation (eqn. 1.10 in the text)

In [13] a more articulated equation was obtained to solve

$$\cos\vartheta_{23} - \cos\vartheta_{21} > 1 - \cos\vartheta_{13} \tag{1.11}$$

that deserves still consideration.
The conclusion of all such elaboration is that Bell's inequality may be violated also in the case of experiments based on mental states during perception and cognition of ambiguous figures.

## 5. Neurophysiological Correlates

Although the (1.10) represents a rough approximation respect to the (1.9), it helps us in identifying the correct psychological and neurphysiological procedure we must follow during experimentation in order to identify Bell's inequality violation.
Let us start with the results the we recently obtained in [7]. In order to perform our Test A and the tests A/B we adopted always the same visual angle. We repeat here that it has been shown that perception and cognition in ambiguous figures is influenced by visual angle [16]. Therefore a constant visual angle $V = 2arctg(S/2d) = 0.33\ rad.$ was used by us in our experimentation with $S$ object's frontal linear size and $d$ distance from the center of the eyes for all the subjects. Let us represent in Figure 3 the neurophysiological condition.

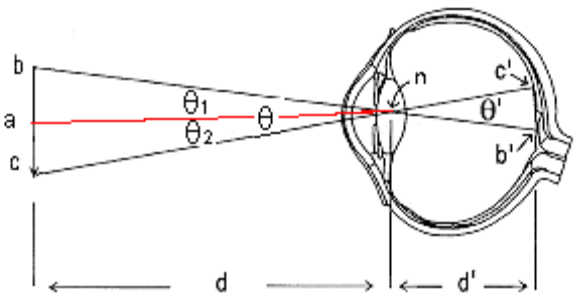

**Figure3**: visual angle and its neurophysiological correlate
through the region of interested retina

We see that, among the possible stimulus variable affecting the reversal phenomenon, we have also the size of the ambiguous figure. The reason is very complicated from the viewpoint of the physiology but in brief we may say that, when the size of the pattern changes, the visual angle subtended from the observer's eye varies and with it varies also the portion of retina involved. A net relation is established among visual angle, quantities describing perceptual alternation and neurophysiological correlates. From the experimental point of view, we must realize figures each time with different side of the figure given as Test (consider a Necker cube to simplify). Each time, during the Test, we will have a different portion of retina involved and with a selected rotation of the retina employed. It will represent the required relation between the neurophysyology and the spinlike quantization rotation that we need. Conditional probabilities will vary each time finally finding those giving Bell's inequality violation.
No doubt may exist that in perspective the quantum mechanics will be strongly improved and powered. According to Orlov [15], this is a theory that reaches a so profound state of elaboration and description that is able to account at the same time of the physical processes of our reality as well as of the perceptive and cognitive features that pertain to our reality. A confirmation of this thesis resides in the fact that we may also write very simple textbooks of quantum mechanics, using as formal instruments only the algebraic set $(e_1, e_2, e_3)$, and developing full arguments as the quantization, the harmonic oscillator, the orbital angular momentum, the hydrogen atom, the time evolution of quantum systems, the EPR and Bell's inequality and other basic foundations of such theory never using the traditional methods of quantum theory but only this algebraic set $(e_1, e_2, e_3)$, and its isomorphic matrices [17]. Quantum mechanics is a Giano Bifronte (two-faced Giano, a mythological God of the past), looking from one hand to physical field and from the other hand to the sphere of mental reality.

## 6. Conclusion

A conclusion seems to appear in its evidence.

A number of statistical tests of quantum like behaviour of cognitive systems has been elaborated and corresponding experiments were designed and performed. In [7, 14] we presented results of our theoretical and experimental research on mental probabilistic interference. In this paper we studied (again both theoretically and experimentally) a Bell-type test for perception of ambiguous figures. This test occurs to be essentially more complicated from the point of view of experimental methodology than our previous test on the interference of probabilities.. We performed detailed analysis (including quantum mechanical probabilistic framework in coupling with neurophysiology) which demonstrated that, although at we have not yet been able to violate Bell's inequality in the ambiguous - perception experiments, there are strong theoretical arguments supporting our expectation to violate it. We explored the method used in quantum mechanics for description of spin-observables. It is well known that algebra of corresponding operators has a direct coupling to geometry – via representing spin observable by trigonometric linear combinations of Pauli matrices. We proceeded by identifying spin geometry with geometry given by the visual angle and its neurophysiological correlate through retina. We hope that our analysis provides the solid ground for further investigations on quantum like behaviour of cognitive systems.